\documentclass{emulateapj}

\begin{document}

\title{Constraints on Fundamental Cosmological Parameters with Upcoming Redshifted 21 cm Observations}

\author{Judd D. Bowman\altaffilmark{1}, Miguel F. Morales\altaffilmark{2}, Jacqueline N. Hewitt\altaffilmark{1} }

\altaffiltext{1}{MIT Kavli Institute for Astrophysics and Space
Research, 77 Massachusetts Avenue, Cambridge, MA 02139, USA}

\altaffiltext{2}{Harvard-Smithsonian Center for Astrophysics, 60
Garden Street, Cambridge, MA 02138, USA}

\email{jdbowman@mit.edu}

\begin{abstract}
Constraints on cosmological parameters from upcoming measurements
with the Mileura Widefield Array---Low Frequency Demonstrator
(MWA-LFD) of the redshifted 21 cm power spectrum are forecast
assuming a flat $\Lambda$CDM cosmology and that the reionization of
neutral hydrogen in the intergalactic medium occurs below a redshift
of $z=8$. We find that observations with the MWA-LFD cannot constrain
the underlying cosmology in this scenario.  In principle, a similar
experiment with a 10-fold increase in collecting area could provide
useful constraints on the slope of the inflationary power spectrum,
$n_{s}$, and the running of the spectral index, $\alpha_{s}$, but
these constraints are subject to the caveat that even a small
reionization contribution could be confused with the cosmological
signal.  In addition to the redshifted 21 cm signal, we include two
nuisance components in our analysis related to the systematics and
astrophysical foregrounds present in low-frequency radio
observations. These components are found to be well separated from
the signal and contribute little uncertainty ($<$30\%) to the
measured values of the cosmological model parameters.
\end{abstract}


\section{INTRODUCTION}

Hydrogen gas in the intergalactic medium (IGM) is a promising tool
for studying some of the most interesting topics in astrophysics,
including the nature of the first luminous objects, the processes of
structure formation, and cosmology. Prior to the epoch of
reionization (EOR), when radiation from the first luminous sources
reionized the IGM, the 21 cm hyperfine line from neutral hydrogen is
predicted to be visible against the cosmic microwave background (CMB)
with relative brightness temperature of order $\pm$10 mK. At high
redshifts, the spin temperature of HI is expected to be less than the
CMB temperature, making the line visible in absorption, whereas at
lower redshifts, after radiation from the first luminous objects has
begun to heat the IGM but before it has reionized it, the line should
be evident as emission.

As the primordial hydrogen cools following recombination and later
reheats, density contrasts are revealed as fluctuations in the
brightness temperature of the redshifted 21 cm line
\citep{1972A&A....20..189S, 1979MNRAS.188..791H, 1990MNRAS.247..510S,
2002ApJ...572L.123I, 2003MNRAS.341...81I, 2004PhRvL..92u1301L,
2005ApJ...626....1B}. At high redshifts, the power spectrum of these
fluctuations is expected to follow closely the dark matter power
spectrum and measurements from this epoch could, independently from
CMB and galaxy clustering experiments, constrain cosmological models.
At lower redshifts, as the first luminous objects ionize their
surroundings, voids are expected to appear in the diffuse emission
\citep{1997ApJ...475..429M, 2000ApJ...528..597T, 2003ApJ...596....1C,
2004ApJ...608..622Z, 2004ApJ...613...16F}. Measurements of the power
spectrum during this epoch would help reveal the characteristics of
the first luminous objects.

Recent efforts have distinguished theoretical aspects of redshifted
21 cm observations that are particularly relevant for constraining
cosmological models. \citet{2005MNRAS.363L..36B} discuss the ability
of redshifted 21 cm observations to constrain the geometry of the
high redshift universe between recombination and reionization;
\citet{2005MNRAS.363..251A} and \citet{2005ApJ...624L..65B} have
shown that differences in the line-of-sight versus angular components
of a redshifted 21 cm power spectrum measurement can be used to
separate primordial density perturbations from features caused by the
radiative processes responsible for reionization; and
\citet{Barkana_AP} has discussed the application of the
Alcock-Paczynski (AP) test to redshifted 21 cm measurements.
Additionally, \citet{2005ApJ...626....1B} consider the effects of the
earliest galaxies on the redshifted 21 cm fluctuations and
\citet{2005MNRAS.362.1047N} discuss using redshifted 21 cm
observations to study the thermal history of hydrogen gas by
detecting a cutoff in the power spectrum due to thermal broadening of
the line.

The experimental challenges of detecting and characterizing neutral
hydrogen in the IGM at high redshift are significant.  While reaching
the sensitivity limit needed to detect the redshifted 21 cm radio
background is readily achievable by very small experiments,
statistical measurements of the fluctuation power spectrum require
the collecting area of the much larger arrays---MWA-LFD, LOFAR, and
PAST---currently under development \citep{2006ApJ...638...20B}. Even
after the required collecting areas have been reached, anticipated
challenges remain. Astrophysical foreground contaminants are five
orders of magnitude brighter than the redshifted 21 cm emission.
Removing the foreground signatures from observations requires
extremely precise observational calibration.

In this paper, we expand on earlier calculations of the sensitivity
of the first generation arrays to forecast their ability to place
constraints on cosmological models from measurements of the
redshifted 21 cm power spectrum.  In order to treat the best-case
scenario and to simplify the analysis, we have chosen to assume that
reionization has not yet begun by the target redshift range of $8 < z
< 10$.  This assumption represents the most optimistic scenario
generally consistent with the existing evidence about when
reionization began from quasar absorption line measurements
\citep{2001ApJ...560L...5D, 2001AJ....122.2850B, 2003AJ....125.1649F,
2004Natur.427..815W} and from Thomson scattering measurements by the
WMAP satellite \citep{2006astro.ph..3449S}.  These measurements
suggest reionization occurred at redshifts $z\gtrsim6$ and $z\sim10$,
respectively.

We begin in Section 2 by reviewing the observational process of
measuring the highly redshifted 21 cm power spectrum and describing
the fiducial experiments.  In Section 3, the method to forecast
constraints on cosmological parameters is described in terms of
statistical errors using a Fisher matrix treatment of the full
three-dimensional power spectrum. The calculation marginalizes over
several cosmological parameters and two anticipated contributions
related to the astrophysical foreground contaminants. The results are
discussed in Section 4 and encapsulated in Figure \ref{f2}.

A similar study has been performed in parallel by
\citet{McQuinn_Cosmology}.  While we analyze the characteristics
relevant to constraining cosmological models with the initial
generation of experiments and in the presence of two nuisance
contributions, their analysis explores the benefit of combining
redshifted 21 cm measurements with information from other
experiments, such as WMAP and Planck. Together, the two efforts
provide a thorough overview of the potential of future redshifted 21
cm power spectrum measurements to contribute to cosmology.

\section{THE MEASUREMENT}

The basis of the statistical measurement of the redshifted 21 cm
power spectrum using low-frequency, wide-field radio observations has
been developed in the literature by \citet{2004ApJ...615....7M},
\citet{2004ApJ...608..622Z}, \citet{2005ApJ...619..678M}, and
\citet{2006ApJ...638...20B}.  These efforts are built on the similar
approach employed for interferometric measurements of CMB
anisotropies \citep{1999ApJ...514...12W, 2002MNRAS.334..569H,
2003ApJ...591..575M} Here, we review the relevant properties of the
measurement; we direct the reader to the previous works for
additional details.

Neutral hydrogen is optically thin to the 21 cm line. Thus, through
the combination of their angular and spectral responses, the
visibility measurements of low-frequency radio arrays inherently
sample the emission from a three-dimensional volume of space at high
redshift \citep{2004ApJ...615....7M}. The measurements may be
represented in a fully Fourier domain and expressed in coordinates
convenient for studying cosmology ($\mathbf{k} \equiv
k_{1},k_{2},k_{3}$) or in coordinates more directly related to
instrumental considerations ($u, v, \eta$).

The fluctuation power spectrum of redshifted 21 cm emission in the
sampled volume of space is mapped to an instrumental response by the
convolution of the power spectrum, $P_{HI}(\mathbf{k})$, with the
instrumental window function, $W(\mathbf{k})$, according to
\citep{2004ApJ...615....7M}
\begin{equation}
C^{HI}(\mathbf{k}) = \left < | \Delta I(\mathbf{k})|^2 \right > =
\int P_{HI}(\mathbf{k}) | W(\mathbf{k}-\mathbf{k}') |^2 d^3k',
\end{equation}
where the window function is given by the Fourier transform of the
instruments angular and frequency response, and is very sharply
peaked for the the cases we will consider (see Section 2.2).

There is an inherent uncertainty in the measurement of the redshifted
21 cm power spectrum from cosmic sample variance.  The uncertainty
can be estimated by dividing the observed three-dimensional Fourier
space into a large number of independent cells, where the volume of
each cell is approximately the size of the window function,
$W(\mathbf{k})$. For the regime of interest in our calculations, the
variance in the measured power spectrum is then
$C_{ij}^{V}(\mathbf{k}) \approx C^{HI}(\mathbf{k})\delta_{ij}$, where
the indices $i$ and $j$ run over all the independent cells in the
sampled volume.

\subsection{Contributions from Astrophysical Foregrounds}
\label{sec_foregrounds}

The redshifted 21 cm emission is not the only source of power in
low-frequency radio observations.  Astrophysical foregrounds are
several orders of magnitude stronger and dominate the measured
signal.  The foregrounds include free-free and synchrotron emission
from the Galaxy, extragalactic point sources, and free-free emission
from elections in the IGM.

The bright foregrounds manifest themselves in several ways.  First,
the Galactic emission dominates the thermal noise in the
measurements, especially at lower frequencies. Although much of this
large-scale emission is expected to be resolved-out by
interferometric observations, its effect on antenna temperature
remains.

The contribution to the measured power spectrum due to thermal noise
is substantial, but white ($\left < C^N(\mathbf{k}) \right > =
constant$), and therefore should be readily removed. In our
calculations, we will assume that it has been removed (but allow for
an imperfect subtraction by including an offset parameter in the
model). After this subtraction, the thermal uncertainty per
independent cell in the three-dimensional measured power spectrum can
be approximated, in instrumental coordinates, using \citep[His Eqn.
11]{2005ApJ...619..678M}
\begin{equation}
\label{eqn_noise} \left [C_{ij}^{N}(\mathbf{u})\right ]_{\rm rms}=
2\left (\frac{ 2 k_{B}T_{sys}}{\epsilon\, dA\,d\eta}\right
)^{2}\frac{1}{B\,\bar{n}(\mathbf{u})\,t}\delta_{ij},
\end{equation}
where $dA$ is the physical antenna area, $d\eta$ is the inverse of
the total bandwidth, $k_B$ is Boltzmann's constant, $T_{sys}$ is the
total system temperature, $\epsilon$ is the efficiency, $B$ is the
total bandwidth, $\overline{n}$ is the time average number of
baselines in an observing cell, and $t$ is the total observation
time.  This equation is derived from the corresponding relationship
for interferometric measurements of CMB anisotropies in \citet[Their
Eqn. 16]{1999ApJ...514...12W}. Although the measured power spectrum
in this case is three-dimensional, the parameters in Equation
\ref{eqn_noise} are approximately independent of frequency within the
observing band. The thermal uncertainty per independent cell,
therefore, is taken also to be independent of $\eta$ (and $k_3$).
Additionally, for the MWA-LFD, the time-averaged visibility
distributions are expected to be dense and to have nearly circular
symmetry, thus $\bar{n}(\mathbf{u})$ depends only on $\sqrt{u^2 +
v^2}$.

The second important effect of astrophysical foregrounds is that they
produce their own signatures in the measured power spectrum.  The
dominate foreground signature is expected to be due to extragalactic
(and Galactic) point sources, which introduces a distinct component
in the observations that is very bright (ranging up to $\sim$1000 K
for individual sources) and highly structured on the sky.
Fortunately, these foreground contaminants are expected to be
distinct from the redshifted 21 cm signal in their angular
correlation functions and in their frequency behavior; and recent
studies indicate that statistical techniques should provide methods
to separate them from the desired signal \citep{2003MNRAS.346..871O,
2004MNRAS.355.1053D, 2004ApJ...608..622Z, 2004ApJ...615....7M,
2005ApJ...625..575S, Wang_Foreground_Subtraction,
2005astro.ph.10027M}.

We will assume that the foregrounds have been cleaned from the
measured signal. Even the best subtraction algorithms leave residual
traces of the contaminate, however, and a perfect model may produce
an imperfect subtraction simply due to thermal uncertainty in the
measurement.  Such a subtraction error would result in a power
spectrum signature following the shapes of the model components
\citep{2005astro.ph.10027M}. The principle residual signature due to
this kind of subtraction error for extragalactic point sources can be
quantified as a $k_3$-dependent factor, $C^F(k_3) \propto k_3^{-2}$.
Since extragalactic point sources are expected to be the most
significant astrophysical foreground contaminate we include this
contribution in our analysis as a nuisance term.

\subsection{Reference Experiments}
\label{sec_ref_exp}
    \begin{figure}
    \begin{center}
    \includegraphics[width = 3.3in]{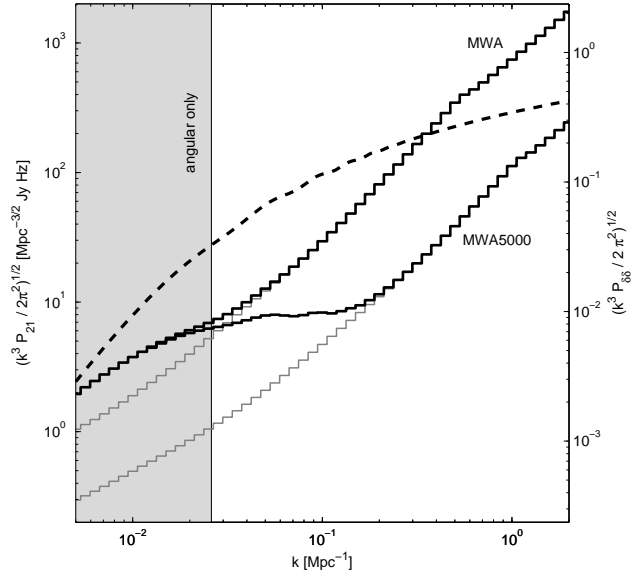}
    \end{center}
    \caption{ \label{f_sensitivity}
Uncertainties (stepped lines) in spherically averaged power spectrum
measurements due to thermal and cosmic sample variance for a single
$\Delta z \simeq 0.5$ (8 MHz) region  at redshift $z=8$ observed by
the MWA and MWA5000 reference experiments relative to the fiducial,
redshifted 21 cm power spectrum (dashed line). The thermal uncertain
(thin gray lines) is overshadowed by uncertainty due to cosmic sample
variance at large scales (small $k$). The combined uncertainty is
shown as black lines.  Due to the limited redshift range over which
cosmic evolution can be neglected ($\Delta z \simeq 0.5$), the
measurements in the light-gray region to the left of the vertical
line are constrained only by angular fluctuations and will thus be
most affected by astrophysical foreground contamination.  The
predicted constraints on the cosmological model parameters in Figure
\ref{f2} and Tables \ref{tab_cov2} and \ref{tab_cov3} were derived
using the full three-dimensional power spectrum, and thus, this plot
of the uncertainty in the spherically averaged measurement is best
used as a guide to illustrate which scale sizes will be most
important in the constraints. }
    \end{figure}
The selection of reference experiments is critical to the outcome of
the forecasting calculations since the uncertainties in redshifted 21
cm power spectrum measurements are coupled very closely to instrument
design.  The first generation of low-frequency radio arrays developed
to perform redshifted 21 cm measurements includes three experiments:
the Low Frequency Array\footnote{http://www.lofar.org} (LOFAR) being
constructed in the Netherlands, the Primeval Structure
Telescope\footnote{\citet{Pen_PAST}} (PAST) under development in
northwestern China, and the
MWA-LFD\footnote{http://haystack.mit.edu/ast/arrays/mwa} in western
Australia. The designs of these instruments reflect different
priorities and different approaches, but they all result in similar
predicted sensitivities for measurements of the redshifted 21 cm
power spectrum.  For example, although LOFAR will have the largest
total collecting area (by a factor of $\sim$10), its planned antenna
distribution scheme is better suited for imaging applications than
power spectrum measurements and its sensitivity is consequently
lessened. The design of the MWA-LFD, on the other hand, has been
optimized for power spectrum measurements and exploits the wide
fields-of-view that are inherent in dipole-based systems to overcome
its more limited collecting area.

Because of its focus on measurements of the redshifted 21 cm power
spectrum, we have chosen to base the design of the reference
experiments in our calculations on the MWA-LFD. We will consider two
experiments. The first is identical to the fiducial MWA-LFD design
described in \citet[Their Section 2]{2006ApJ...638...20B}. The array
design consists of $N=500$ antennas distributed within a $D=1500$ m
diameter circle. The density of antennas as a function of radius is
taken to go as $\sim r^{-2}$, but capped at a maximum density of one
antenna per 18 m$^2$. The antenna response is approximated by
\begin{equation}
    \begin{array}{cc}
        W(\theta) = \cos^2 \left ( \frac{\pi}{2} \theta / \Theta \right ), & \quad \mbox{ $\theta < \Theta$} \\
    \end{array}
\end{equation}
where $\Theta$ is proportional to wavelength and spans from 31 to
38$^\circ$ over the range of interest. The angular resolution of the
array is given by $\lambda/D$ and the total collecting area by
$N~dA$, where $dA$ is the collecting area of each antenna and scales
like $dA = 16 ( \lambda^2/4 )$ for $\lambda < 2.1$ m and is capped
for longer wavelengths. Finally, the full bandwidth of the instrument
is $B=32$ MHz and the frequency resolution is $8$ kHz. All of the
fiducial properties are summarized in Tables \ref{tab_fiducial} and
\ref{tab_redshift}.

For the analysis in this paper, we define the observation to be of a
single field with 1000 hours integration during the most favorable
circumstances. Additionally, we set the frequency coverage to
$125<f<157$ MHz, which spans $10 > z > 8$, and treat the bandwidth as
four consecutive $8$ MHz regions, each of which spans approximately
$\Delta z = 0.5$.  In principle, the observed 21 cm power spectrum is
varying continuously with redshift due to cosmology-dependent
effects.  By dividing our measured volume of space into thin regions
in redshift we may ignore cosmic evolution within each region.

The second reference experiment is based on an expanded MWA-LFD
configuration similar to the one considered by
\citet{2005ApJ...634..715W}, although more condensed. Dubbed the
MWA5000, the expanded array contains 5000 antennas distributed over a
3000 m diameter region. The density of antennas remains $\sim r^{-2}$
and capped at one per 18 m$^2$. The antenna response, instrument
bandwidth, and frequency resolution are also unchanged (see Tables
\ref{tab_fiducial} and \ref{tab_redshift}). In both experiments all
antenna elements are correlated to preserve the large field of view.

Figure \ref{f_sensitivity} illustrates the relative sensitivities of
the two reference experiments by plotting their uncertainties in
spherically averaged bins due to thermal noise and cosmic sample
variance along with a fiducial, redshifted 21 cm power spectrum.  The
two reference experiments share a common level of uncertainty at
large spatial scales (small $k$) due to cosmic sample variance since
their fields of view are identical. At smaller spatial scales
($k\gtrsim0.1$), however, the larger collecting area of the MWA5000
reduces its thermal uncertainty in a given integration time compared
to that of the MWA. The MWA5000 is nearly an order of magnitude more
sensitivity than the MWA at these scales.

\section{METHOD}

The Fisher information matrix provides a convenient method for
translating uncertainties in power spectrum measurements to
constraints on cosmological parameters \citep{1997ApJ...480...22T}.
The minimum errors are calculated using the Fisher matrix:
\begin{equation}
F_{ab} = \sum_{i} \frac{1}{\sigma(\mathbf{k}_i)^2} \frac{\partial
P(\mathbf{k}_i)}{\partial p_a} \frac{\partial
P(\mathbf{k}_i)}{\partial p_b},
\end{equation}
where $\sigma^2 \equiv \sigma_N^2 + \sigma_V^2$ is the combined
uncertainty per independent cell due to thermal noise and cosmic
sample variance, $P$ is the measured power in a cell given by
Equation \ref{eqn_p_all}, and $p$ is the set of model parameters.
Taking the square roots of the diagonal elements of the inverse of
Fisher matrix gives the errors.

The model we use for the observed power spectrum represents the
measurement after substantial data reduction has occurred, including
instrumental calibration and astrophysical foreground subtraction. At
this stage, the observed power spectrum is parameterized based on
three contributions:
\begin{equation}
P(\mathbf{k}) \equiv P_{HI}(k) + P_{F}(k_3) + P_{N},
\label{eqn_p_all}
\end{equation}
where the first term is the redshifted 21 cm contribution, the second
is the primary residual astrophysical (extragalactic point source)
foreground contribution discussed in Section 2.1, and the third is
the residual white thermal noise contribution (also discussed in
Section 2.1).

For our assumption that reionization has heated the neutral hydrogen
in the IGM but not yet produced significant features in the emission
patterns of the gas by the target redshifts, the redshifted 21 cm
contribution to the model power spectrum follows the matter power
spectrum and can be expressed at a given redshift as
\begin{equation}
P_{HI}(k) \equiv C_{Jy}^2 P_{\delta \delta}(k \; | A, \Omega_M,
\Omega_b, h, n_s, \alpha_s), \label{eqn_p_hi}
\end{equation}
where $C_{Jy}$ is the strength of the 21 cm emission from
mean-density neutral hydrogen gas in the IGM and $P_{\delta
\delta}(\mathbf{k})$ is the matter power spectrum. At the target
redshifts, fluctuations in the matter density are expected generally
to be in the linear regime over the range of spatial scales sampled.
For redshift $z=8$, the perturbations at the smallest scales
constrained by the measurements ($k\simeq1$ Mpc$^{-1}$) will be
approximately 40\%; and at the largest scales ($k\simeq0.01$
Mpc$^{-1}$), the perturbations will be less than 1\%. In contrast to
measurements of the matter power spectrum through large scale
structure surveys of galaxies, redshifted 21 cm measurements probe
directly the baryon density perturbations in the IGM, which follow
closely the dark matter density perturbations. The effect of small
deviations from the purely linear regime can be modeled easily in
this case and we omit, therefore, this complication in our analysis.
We use six parameters, $p=\{A, \Omega_M, \Omega_b, h, n_s,
\alpha_s\}$, to describe the matter power spectrum. These parameters
are summarized in Table \ref{tab_cosmo} along with their fiducial
values. Additionally, we constrain $\Omega_K=0$ and $\Omega_\Lambda =
1 - \Omega_M$.  To compute the matter power spectrum, we use CMBFAST
\citep{1996ApJ...469..437S} and do not include velocity distortions
or deviations from linear dynamics in the model.


As described in Section 2.1, the primary residual foreground
contribution is a function of $k_3$ and is expressed as
\begin{equation}
P_{F}(k_3) \equiv F \times c_{norm} \left ( \frac{k_3}{k_{norm}}
\right )^{-2}, \label{eqn_p_f}
\end{equation}
where $F$ is an amplitude scale factor of order unity and $c_{norm}$
is a normalization constant. The thermal offset contribution does not
vary with $\mathbf{k}$ and is defined as
\begin{equation}
P_N\equiv~N~\times~c_{norm},
\end{equation}
where $N$ is also an amplitude scale factor of order unity. For the
Fisher matrix analysis, both $P_F$ and $P_N$ are normalized by
setting $k_{norm} = 0.01$~Mpc$^{-1}$ and evaluating $c_{norm} =
P_{HI}(k_{norm})$ at redshift $z=8$. Thus, the magnitudes of $F$ and
$N$ give the amplitudes of the residual contributions approximately
relative to the peak value of the redshifted 21 cm power spectrum.
Adding $F$ and $N$ to the set of six cosmological parameters yields
the eight parameters, $p=\{A, \Omega_M, \Omega_b, h, n_s, \alpha_s,
F, N\}$, which constitute the complete set of free variables of our
base model.

\section{RESULTS}
    \begin{figure*}
    \begin{center}
    \plotone{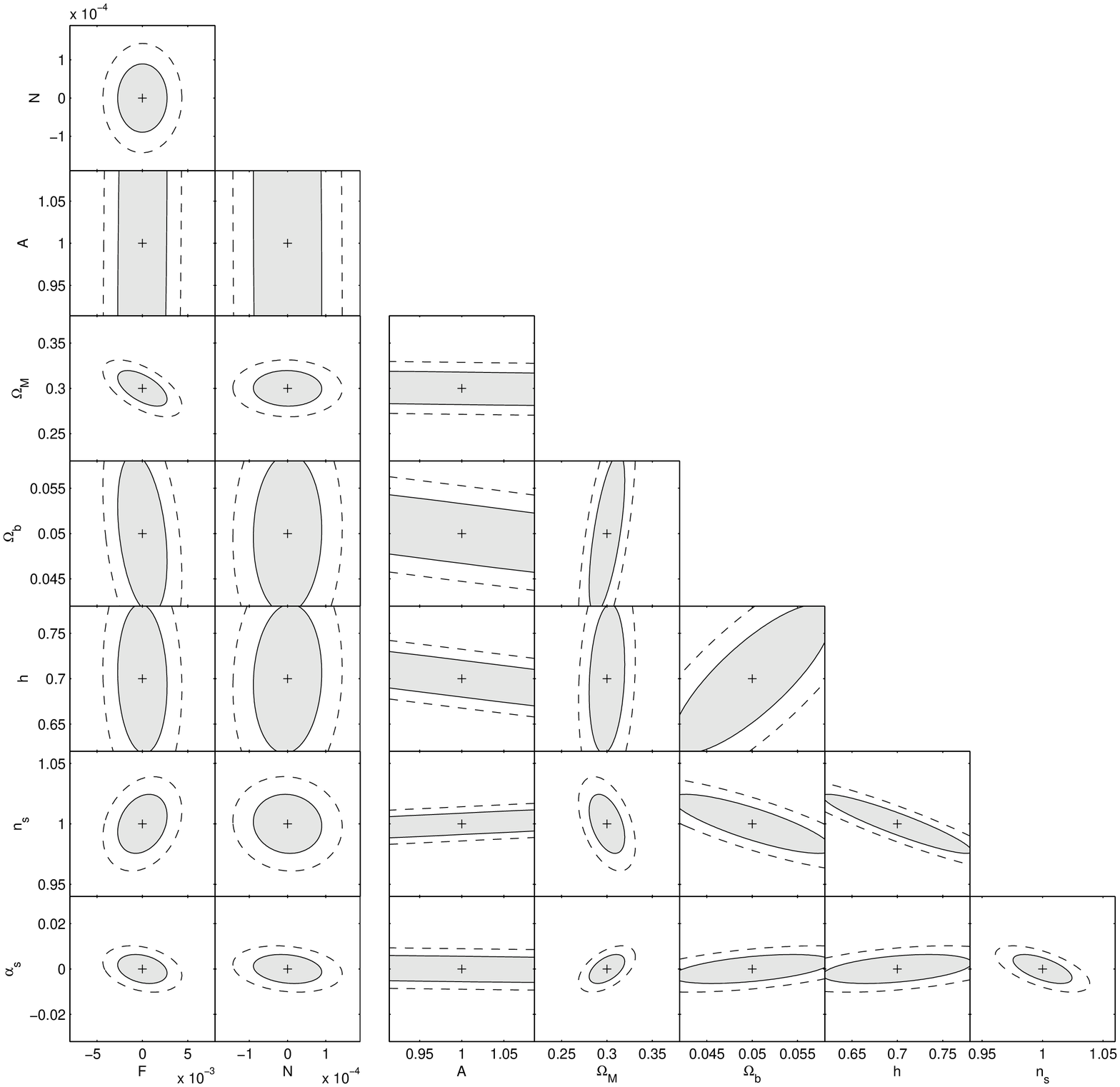}
    \end{center}
    \caption{\label{f2}
    Marginalized elliptical error regions for pairs of model parameters
    for the MWA5000 reference experiment. The contours are for 68\%
    (solid) and 95\% (dashed) likelihood. The first two columns are set
    apart to emphasize the distinction between the nuisance components
    in the model due to astrophysical foregrounds and the cosmological terms.
    The bounds of the plotted regions for the cosmological parameters are
    set at twice the uncertainties ($\pm$ 2-$\sigma$) reported in the
    WMAP first-year results for the respective parameters.  Thus, if the
    dotted ellipse is visible in a particular plot, the constraints on
    the parameters would be an improvement over WMAP.  It is
    evident that the MWA5000 would do a relatively good job
    of constraining $\Omega_M$, $n_s$, and $\alpha_s$, and a
    comparatively poor job of constraining $\Omega_b$ and $h$.  The
    inability of the MWA5000 to constrain the scalar amplitude, A, of the
    power spectrum is due to degeneracies with other parameters, in
    particular $\Omega_b$ and $h$, but also $n_s$.
    }
    \end{figure*}
The results of performing the Fisher matrix calculations for the two
reference experiments described in Section \ref{sec_ref_exp} are
summarized in Table \ref{tab_cov1}, which lists the forecasted
1-$\sigma$ uncertainties on the model parameters. These values
indicate that observations with the MWA-LFD would not constrain the
cosmological parameters significantly. The only exception to this
finding is for $\alpha_s$, which could be constrained to $0\pm0.04$
under the favorable assumptions used in the model. The second
reference experiment, MWA5000, does constrain reasonably all the
model parameters.  Table \ref{tab_cov1} shows that this reference
experiment provides constraints at levels approximately equivalent to
those from the first-year WMAP results \citep[Their Table
10]{2003ApJS..148..175S}.

The elements of the covariance matrices for each reference experiment
are given in Tables \ref{tab_cov2} and \ref{tab_cov3}, and additional
information about the forecasted constraints for the MWA5000 is
provided in Figure \ref{f2}, which shows marginalized error ellipses
for two-parameter combinations of the model parameters.  The
properties of these results are discussed in detail in the remainder
of this section.

\subsection{Parameter Dependencies}
    \begin{figure}
    \begin{center}
    \includegraphics[width = 3.3in]{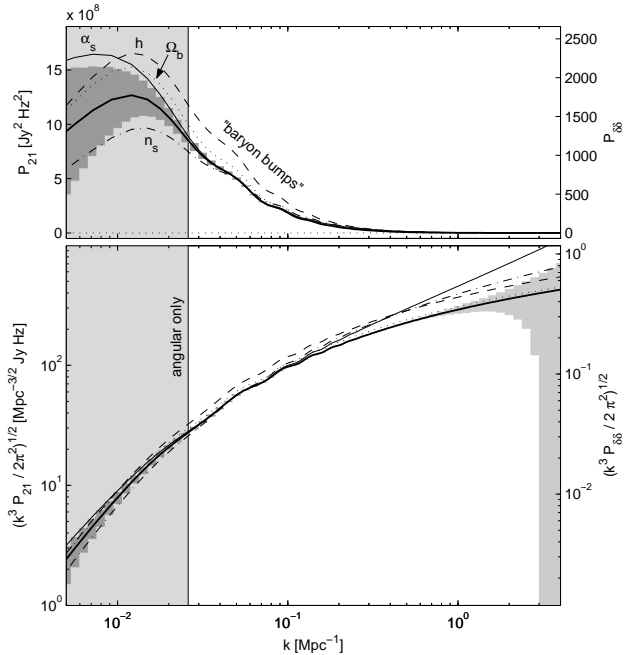}
    \end{center}
    \caption{ \label{f1}
Redshifted 21 cm power spectrum for the fiducial cosmological model
parameters, shown with four example variations of the parameter
values. The top and bottom panels show the same elements plotted in
different units. The thick solid line is for the fiducial model with
standard cosmological parameters ($\Omega_b=0.05$, $\Omega_M=0.30$,
$h=0.70$, $n_s=1.00$, $\alpha_s=0$).  The other curves are produced
by adjusting one cosmological parameter from the fiducial: $\Omega_b$
+ 10\% (dotted), $h$ + 10\% (dash), $n_s$ + 10\% (dash dot), and
$\alpha_s=0.1$ (thin solid).  The dark shaded region indicates the
1-$\sigma$ uncertainty due to cosmic variance within the observed
field and the light shaded region (at large $k$) is the thermal
uncertainty for the MWA5000 reference experiment. As in Figure
\ref{f_sensitivity}, the measurements to the left of the vertical
line are constrained by angular fluctuations only and will be most
sensitive to astrophysical foreground contamination.}
    \end{figure}
To facilitate a conceptual understanding of the origin of the
constraints on the model parameters, it is instructive to
individually vary the cosmological parameters and compare the
resulting redshifted 21 cm signals. On all but the largest length
scales, the experimentally observed power spectrum is well
approximated by
\begin{equation}
\label{eqn_p_hi2}
P_{\rm HI}(u,v,\eta) \cong W^2 C_{\rm Jy}^2 \left \{ P_{\delta
\delta}(\mathbf{ k}) \frac{d^3\mathbf{ k}}{dudvd\eta} \right \},
\end{equation}
where $W$ is the integrated value of the observational window
function and represents the signal strength for a given instrument,
and $d^3\mathbf{ k}/dudvd\eta $ is the Jacobian for converting the
units of the matter power spectrum to the observed units.  The
relationships between $u,v,\eta$ and $\mathbf{k}$ are
\citep{2004ApJ...615....7M}:
    \begin{eqnarray}
    [k_1, k_2] = \frac{ 2 \pi }{ D_M(z) } [u,v] \label{kpara}\\
    k_3 \approx \frac{ 2 \pi \nu_{21} H_0  E(z) }{ c \left (1 + z \right)2 }
    \eta, \label{kperp}
    \end{eqnarray}
where
    \begin{equation}
    E(z) \equiv \sqrt{ \Omega_M (1+z)^3 + \Omega_k (1+z)^2 + \Omega_\Lambda
    (1+z)^{3(1+w)}},
    \end{equation}
and $D_M(z)$ and $\nu_{21} = 1420.4$ MHz are the transverse comoving
distance and the rest-frame frequency of the hyperfine line,
respectively.

In Equation \ref{eqn_p_hi2}, there are three separate contributions
which may be affected by changing the cosmological parameters: the
matter power spectrum itself, the neutral hydrogen emission factor,
and the coordinate mapping between $\mathbf{k}$ and $u,v,\eta$. The
mapping between $\mathbf{k}$ and u,v,$\eta$ produces two effects when
measuring the power spectrum.  First, the region in $u,v$ and $\eta$
occupied by the observed power spectrum depends on the cosmological
model.  This effect is analogous to the observed CMB power spectrum
shifting in $\ell$ for different cosmological models, but since the
redshifted 21 signal is three-dimensional, there may be a difference
in scaling between the sky-plane and line-of-sight dimensions. This
additional degree of freedom gives rise to the AP test
\citep{1979Natur.281..358A} (see also \citet{2003ApJ...598..720S} for
an applicable discussion of constraining dark energy using baryon
acoustic oscillations in power spectra from three-dimensional galaxy
redshift surveys).

The second effect of coordinate mapping in the observed power
spectrum is the presence of a significant cosmology-dependent
amplitude factor in the measured power spectrum due to the Jacobian
contribution in Equation \ref{eqn_p_hi2}. Differentiating and
combining Equations \ref{kpara} and \ref{kperp} yields the Jacobian:
\begin{equation}
\label{jacobian} \frac{d^3\mathbf{ k}}{du dv d\eta}  = \frac{ \left
(2 \pi \right )^3 \nu_{21} H_0  E(z) }{ c \left ( 1 + z \right)^2
D_M^2(z)}.
\end{equation}

The effect of the Jacobian amplitude factor is further compounded by
the hydrogen emission factor, $C_{Jy}$. This term is unique to
redshifted 21 cm experiments and provides a significant boost to the
cosmology-dependent amplitude factor. Assuming the spin temperature
of the neutral hydrogen in the IGM is significantly warmer than the
CMB, the hydrogen emission factor goes as \citep[Their Equation 13
and Their Appendix A]{2004ApJ...615....7M}
\begin{equation}
\label{cyeq} C_{Jy} \propto \frac{\Omega_b h}{E(z)} \bar{x}_{\rm HI},
\end{equation}
where $\bar{x}_{\rm HI}$ is the mean neutral fraction (taken to be
unity in this paper).  Using Equations \ref{jacobian} and \ref{cyeq}
and recalling that $D_M(z)\propto h^{-1}$, the amplitude of the
observed power spectrum is now seen to be proportional to
\begin{equation}
\label{amp_eq} P_{HI} \propto \frac{\Omega_b^2 h^5}{ (1+z)^2 E(z)}
P_{\delta \delta}.
\end{equation}
Equation \ref{amp_eq} indicates that the amplitude of the redshifted
21 cm fluctuations does contain useful information about the
cosmological model (with the caveat that the HI spin temperature must
be known). This is in contrast to measurements of the matter power
spectrum by galaxy clustering surveys, where the connection between
the observed amplitude of the power spectrum and the underlying
physical amplitude remains weakly understood.

Figure \ref{f1} shows the spherically-binned redshifted 21 cm power
spectrum signal for the fiducial model (thick line, $\Omega_b=0.05$,
$\Omega_M=0.30$, $h=0.70$, $n_s=1.00$, $\alpha_s=0$) along with the
expected signals obtained by varying individual parameters.  We can
see that varying $\Omega_b$ or $h$ primarily results primarily in a
simple scaling of the observed power spectrum, as expected from
Equation \ref{amp_eq}. Thus these parameters are highly degenerate
with each other and with the amplitude factor A, as shown in Figure
\ref{f2} and Table \ref{tab_cov3}.

The primordial power spectrum slope $n_{s}$ and running of the
spectral index $\alpha_{s}$ do not enter into the overall amplitude
factor in Equation \ref{amp_eq} for the observed power spectrum, but
instead affect the shape of the underlying matter power spectrum.
Increasing $n_{s}$ lowers the power at small $k$ and increases the
the power at large $k$, whereas increasing $\alpha_{s}$ boosts the
power spectrum at both extremes.  Because these parameters do not
strongly affect the amplitude normalization, they are relatively
independent of A, $\Omega_b$, and $h$ as shown in Figure \ref{f2} and
Table \ref{tab_cov3}. They are also independent of our foreground
parameter, $F$, because the shape they introduce is different than
the $k_{3}^{-2}$ shape of the residual foreground contamination. Thus
$n_{s}$ and $\alpha_{s}$ are the best constrained cosmological
parameters for the planned redshifted 21 cm observations.

The last class of observational effects due to the cosmological model
is from the coordinate mapping. Most significantly, for an increase
in $h$, a radially inward shift of scales (see Equations \ref{kpara}
and \ref{kperp}, with $D_M(z) \propto h^{-1}$) counteracts the
amplitude increase from the $C_{Jy}$ and Jacobian factors over much
of the observed range in $k$-space. Increasing $\Omega_M$ also shifts
scales inward (through $E(z)$ and $D_M(z)$), but does so differently
for the line-of-sight direction and the transverse directions. This
AP effect cannot be represented easily in Figure \ref{f1}, however,
we have included it in all the constraints. The significance of the
effect is limited since we take $\Omega_\Lambda = 1 - \Omega_M$. A
10\% increase in $\Omega_M$ yields an approximately 4\% reduction in
scale, but only about 1\% distortion between the directions. In
general, coordinate based effects may be better constrained with
observations targeting the quasar bubbles during reionization since
sharper spatial features in the power spectrum may emerge during
reionization due to characteristic sizes of reionized regions and
Stromgren spheres around quasars.

\subsection{Degeneracies in Amplitude}

The uncertainty in $A$ is considerable. Table \ref{tab_cov1} shows
that the MWA5000 only constrains A to within 45\%.  The origin of
this uncertainty is explained by the amplitude factor discussed above
and is clearly evident in Table \ref{tab_cov3}, which displays the
large degeneracies (as normalized covariance factors with magnitudes
close to unity) between $A$, $\Omega_b$, $h$, and more unexpectedly,
$n_s$. The degeneracy with $n_s$ is due to the range of scales to
which the experiments are sensitive, which is primarily after the
pivot point for the primordial power spectrum spectral index
($k=0.05$ Mpc$^{-1}$). Therefore, changes in $n_s$ tend to appear
less similar to tilting to the power spectrum and more like changing
the amplitude, although $n_s$ remains well constrained. Additionally,
the covariance between $A$ and $\Omega_b$ confirms that the baryon
acoustic oscillations in the power spectrum are not contributing much
information about the baryon density relative to the amplitude
changes from the $C_{Jy}$ and Jacobian factors.

\subsection{Residual Foreground Contamination}

\begin{figure}
\begin{center}
\includegraphics[width = 3.3in]{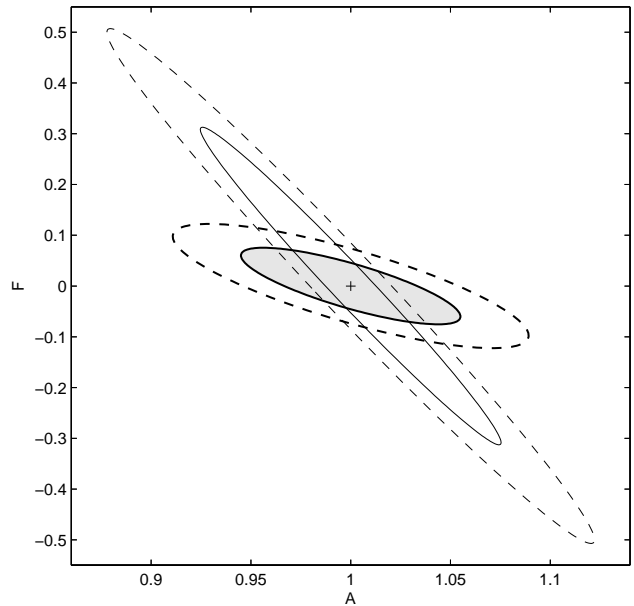}
\end{center}
\caption{ \label{f3} Unmarginalized error ellipses for the amplitudes
of the redshifted 21 cm power spectrum and primary residual
foreground contamination, A and F, forecasted for the MWA-LFD. The
thick curves are for a Fisher matrix calculation utilizing the full
three-dimensional measured power spectrum, while the thin curves are
for the same calculation using spherical bins in $k$-space.  The
contours are for 68\% and 95\% likelihood.  The behavior is similar
for the MWA5000, although the errors are smaller.}
\end{figure}

Mitigating astrophysical foreground contamination is an important
issue for redshifted 21 cm experiments, whether attempting to
constrain cosmological models or determine reionization processes and
scenarios. To this end, the relationships between $P_{HI}$, $P_F$,
and $P_N$, as described by the covariance between our model
parameters A, F, and N are very useful to study in some detail.

The constraints on the amplitudes of the residual astrophysical
foreground and residual thermal offset terms are especially
encouraging. Neither contribution is significantly degenerate with
changes to the redshifted 21 cm power spectrum. In particular, the
long, diminishing tail of the redshifted 21 cm power spectrum
provides many statistical samples which contribute little to
knowledge of the cosmological parameters, yet provide good references
to prevent degeneracy with changes in the magnitude of the residual
thermal offset. Thus, $P_N$ is constrained to better than 2\% of the
peak value of $P_{HI}$ for the MWA-LFD and to better than 0.01\% for
the MWA5000.

As discussed in Section \ref{sec_foregrounds}, the residual
foreground contamination is expected to be a power-law along the
line-of-sight direction and to have little structure in the
transverse directions. This is distinct from the redshifted 21 cm
power spectrum with its generally spherical symmetry, although
geometrical effects due to the shape of the observed volume of space
work to negate this distinction. The observed region has dimensions
of $\sim$9000 by $\sim$9000 Mpc in the sky-plane, but only $\sim$100
Mpc along the line-of-sight \citep{2006ApJ...638...20B}. Since all
three dimensions are sampled with an approximately equal number of
divisions, this flattened shape is inverted when transforming into
$k$-space, leading to a highly elongated data cube in the
$k_3$-direction. For the MWA-LFD, the maximum $k$ in the sky-plane
($k_1$,$k_2$) is approximately $0.05$ Mpc$^{-1}$, while along $k_3$
it is over $10$ Mpc$^{-1}$. Therefore, much of the information in the
measurement about the redshifted 21 cm power spectrum is actually
coming from the line-of-sight direction. Furthermore, the lack of
depth in the observed volume prevents the experiments from probing
large scales in the line-of-sight direction and, thus, the peak of
the power spectrum (at $k \approx 0.01$) is not well constrained.
This means that not only is most of the information coming from the
line-of-sight direction, it is coming from the power law-like tail
($\sim k^{-3}$) of the power spectrum. These effects combine to make
the redshifted 21 cm contribution and residual foreground
contamination less distinguishable in the measured power spectrum.

Fortunately, the analysis indicates that the residual foreground
component is well separated from the redshifted 21 cm power spectrum
and does not affect substantially the ability to constrain the
cosmological parameters. From Table \ref{tab_cov3} and Figure
\ref{f2}, we see that the largest degeneracy is between $F$ and
$\Omega_M$. Removing $P_F$ from the model power spectrum, and thus
$F$ from the parameter set such that $p=$\{$A$, $\Omega_M$,
$\Omega_b$, $h$, $n_s$, $\alpha_s$, $N$\}, would reduce the
forecasted uncertainty on $\Omega_M$ by less than 30\% in an
observation with with MWA5000. The contributions from the residual
thermal offset, $P_N$, are even less.

In our analysis so far we have treated the full three-dimensional
measured power spectrum.  This has maximized the effects of the
symmetry differences between the 21 cm power spectrum and primary
residual foreground power spectrum. We might have instead considered
a reduced one-dimensional power spectrum produced by averaging over
spherical bins in $k$-space.  This approach is used commonly to
reduce data in which the expected signal has approximately spherical
symmetry.

In Figure \ref{f3}, we compare the full three-dimensional treatment
with the spherically-binned approach by plotting unmarginalized error
ellipses in the A-F plane of parameter space for both cases.  It is
clear that the binned method produces a significantly larger
degeneracy between the magnitude of the residual foreground
contribution and the amplitude of the 21 cm neutral hydrogen power
spectrum than the full three-dimensional treatment.

\subsection{Dark Energy Equation of State}

Determining the nature of dark energy has become an important goal in
astrophysics and the applicability of future large-scale structure
surveys to this topic has been considered recently with generally
favorable results \citep{2003ApJ...598..720S, 2003PhRvD..68h3504L,
2003PhRvD..68f3004H, 2004PhRvD..70l3008W, 2005MNRAS.360...27A}. With
their similarities to large-scale structure surveys, would
measurements of the redshifted 21 cm neutral hydrogen power spectrum
also improve knowledge of dark energy?

Including the dark energy equation of state parameter, w, as a free
variable in the Fisher matrix analysis, so that $p = \{A, \Omega_M,
\Omega_b, h, n_s, \alpha_s, w, F, N\}$, yields little information
about the nature of dark energy. Neither the MWA-LFD nor the MWA5000
are able to provide meaningful constraints on $w$ without priors from
other experiments. Furthermore, the constraints on $\Omega_M$,
$\Omega_b$, and $h$ are significantly weakened (by up to an order of
magnitude), although information is retained about the primordial
power spectrum.

Although the MWA5000 reference experiment would be able to detect the
baryon acoustic oscillations (see Figure \ref{f1}), the experiment
targets emission from high redshifts prior to the epoch of dark
energy domination.  The value of the experiment would be relevant
only as an in intermediate measurement between the CMB and low
redshift large-scale structure surveys \citep{Barkana_AP}, but could
be important if dark energy had unusual behavior at high redshifts.

\section{CONCLUSION}

The first generation of redshifted 21 cm experiments have been shown
previously to have the potential to characterize the processes and
history of reionization at redshifts $6<z<12$. In this effort, we
have considered a scenario in which reionization is not the dominant
contribution to the measured power spectrum during this period.  Our
findings demonstrate that the initial experiments would not
contribute substantially to knowledge of the underlying cosmology
under these circumstances.  While they may provide some use to
cosmological studies, the primary science return of the first
generation of redshifted 21 cm experiments will be regarding the
astrophysics of reionization.

Under the optimistic assumptions used in this paper, the
second-generation experiments, similar to the MWA5000 reference
experiment, will be better suited to constrain $\Omega_M$ and the
primordial power spectrum through $n_s$ and $\alpha_s$.  Even for the
spectral index properties it is difficult to envision a scenario
where redshifted 21 cm observations alone could provide an
unambiguous constraint on cosmological information, however, since
the signal could easily be confused if the universe had a small
amount of reionization structure on small scales or spin temperature
fluctuations in the IGM.

We have also considered the covariance between the redshifted 21 cm
power spectrum due to a fully neutral IGM and the primary residual
foregrounds of low-frequency radio observations. The signal and
contamination components were easily distinguished in the full
three-dimensional power spectrum analysis.

\acknowledgements

We would like to thank Matt McQuinn, Matias Zaldarriaga, Oliver Zahn,
and Lloyd Knox for fruitful discussions. Support for this work was
provided by NSF grant AST-0121164 and the MIT School of Science.

%


\begin{thebibliography}{}

\bibitem[\protect\citeauthoryear{{Abdalla} \& {Rawlings}}{{Abdalla} \&
  {Rawlings}}{2005}]{2005MNRAS.360...27A}
{Abdalla}, F.~B.,  \& {Rawlings}, S. 2005, \mnras, 360, 27

\bibitem[\protect\citeauthoryear{{Alcock} \& {Paczynski}}{{Alcock} \&
  {Paczynski}}{1979}]{1979Natur.281..358A}
{Alcock}, C.,  \& {Paczynski}, B. 1979, \nat, 281, 358

\bibitem[\protect\citeauthoryear{{Ali}, {Bharadwaj}, \& {Pandey}}{{Ali}
  et~al.}{2005}]{2005MNRAS.363..251A}
{Ali}, S.~S., {Bharadwaj}, S.,  \& {Pandey}, B. 2005, \mnras, 363, 251

\bibitem[\protect\citeauthoryear{{Barkana}}{{Barkana}}{2006}]{Barkana_AP}
{Barkana}, R. 2006, \mnras, submitted

\bibitem[\protect\citeauthoryear{{Barkana} \& {Loeb}}{{Barkana} \&
  {Loeb}}{2005a}]{2005ApJ...624L..65B}
{Barkana}, R.,  \& {Loeb}, A. 2005a, \apjl, 624, L65

\bibitem[\protect\citeauthoryear{{Barkana} \& {Loeb}}{{Barkana} \&
  {Loeb}}{2005b}]{2005ApJ...626....1B}
{Barkana}, R.,  \& {Loeb}, A. 2005b, \apj, 626, 1

\bibitem[\protect\citeauthoryear{{Barkana} \& {Loeb}}{{Barkana} \&
  {Loeb}}{2005c}]{2005MNRAS.363L..36B}
{Barkana}, R.,  \& {Loeb}, A. 2005c, \mnras, 363, L36

\bibitem[\protect\citeauthoryear{{Becker} et~al.}{{Becker}
  et~al.}{2001}]{2001AJ....122.2850B}
{Becker}, R.~H., et~al. 2001, \aj, 122, 2850

\bibitem[\protect\citeauthoryear{{Bowman}, {Morales}, \& {Hewitt}}{{Bowman}
  et~al.}{2006}]{2006ApJ...638...20B}
{Bowman}, J.~D., {Morales}, M.~F.,  \& {Hewitt}, J.~N. 2006, \apj, 638, 20

\bibitem[\protect\citeauthoryear{{Ciardi} \& {Madau}}{{Ciardi} \&
  {Madau}}{2003}]{2003ApJ...596....1C}
{Ciardi}, B.,  \& {Madau}, P. 2003, \apj, 596, 1

\bibitem[\protect\citeauthoryear{{Di Matteo}, {Ciardi}, \& {Miniati}}{{Di
  Matteo} et~al.}{2004}]{2004MNRAS.355.1053D}
{Di Matteo}, T., {Ciardi}, B.,  \& {Miniati}, F. 2004, \mnras, 355, 1053

\bibitem[\protect\citeauthoryear{{Djorgovski} et~al.}{{Djorgovski}
  et~al.}{2001}]{2001ApJ...560L...5D}
{Djorgovski}, S.~G., {Castro}, S., {Stern}, D.,  \& {Mahabal}, A.~A. 2001,
  \apjl, 560, L5

\bibitem[\protect\citeauthoryear{{Fan} et~al.}{{Fan}
  et~al.}{2003}]{2003AJ....125.1649F}
{Fan}, X., et~al. 2003, \aj, 125, 1649

\bibitem[\protect\citeauthoryear{{Furlanetto}, {Zaldarriaga}, \&
  {Hernquist}}{{Furlanetto} et~al.}{2004}]{2004ApJ...613...16F}
{Furlanetto}, S.~R., {Zaldarriaga}, M.,  \& {Hernquist}, L. 2004, \apj, 613, 16

\bibitem[\protect\citeauthoryear{{Hobson} \& {Maisinger}}{{Hobson} \&
  {Maisinger}}{2002}]{2002MNRAS.334..569H}
{Hobson}, M.~P.,  \& {Maisinger}, K. 2002, \mnras, 334, 569

\bibitem[\protect\citeauthoryear{{Hogan} \& {Rees}}{{Hogan} \&
  {Rees}}{1979}]{1979MNRAS.188..791H}
{Hogan}, C.~J.,  \& {Rees}, M.~J. 1979, \mnras, 188, 791

\bibitem[\protect\citeauthoryear{{Hu} \& {Haiman}}{{Hu} \&
  {Haiman}}{2003}]{2003PhRvD..68f3004H}
{Hu}, W.,  \& {Haiman}, Z. 2003, \prd, 68, 063004

\bibitem[\protect\citeauthoryear{{Iliev} et~al.}{{Iliev}
  et~al.}{2003}]{2003MNRAS.341...81I}
{Iliev}, I.~T., {Scannapieco}, E., {Martel}, H.,  \& {Shapiro}, P.~R. 2003,
  \mnras, 341, 81

\bibitem[\protect\citeauthoryear{{Iliev} et~al.}{{Iliev}
  et~al.}{2002}]{2002ApJ...572L.123I}
{Iliev}, I.~T., {Shapiro}, P.~R., {Ferrara}, A.,  \& {Martel}, H. 2002, \apjl,
  572, L123

\bibitem[\protect\citeauthoryear{{Linder}}{{Linder}}{2003}]{2003PhRvD..68h3504%
L}
{Linder}, E.~V. 2003, \prd, 68, 083504

\bibitem[\protect\citeauthoryear{{Loeb} \& {Zaldarriaga}}{{Loeb} \&
  {Zaldarriaga}}{2004}]{2004PhRvL..92u1301L}
{Loeb}, A.,  \& {Zaldarriaga}, M. 2004, Physical Review Letters, 92, 211301

\bibitem[\protect\citeauthoryear{{Madau}, {Meiksin}, \& {Rees}}{{Madau}
  et~al.}{1997}]{1997ApJ...475..429M}
{Madau}, P., {Meiksin}, A.,  \& {Rees}, M.~J. 1997, \apj, 475, 429

\bibitem[\protect\citeauthoryear{{McQuinn} et~al.}{{McQuinn}
  et~al.}{2006}]{McQuinn_Cosmology}
{McQuinn}, M., {Zahn}, O., {Zaldarriaga}, Z., {Hernquist}, L.,  \&
  {Furlanetto}, S. 2006, \apj, submitted

\bibitem[\protect\citeauthoryear{{Morales}}{{Morales}}{2005}]{2005ApJ...619..6%
78M}
{Morales}, M.~F. 2005, \apj, 619, 678

\bibitem[\protect\citeauthoryear{{Morales}, {Bowman}, \& {Hewitt}}{{Morales}
  et~al.}{2005}]{2005astro.ph.10027M}
{Morales}, M.~F., {Bowman}, J.~D.,  \& {Hewitt}, J.~N. 2005, ArXiv Astrophysics
  e-prints

\bibitem[\protect\citeauthoryear{{Morales} \& {Hewitt}}{{Morales} \&
  {Hewitt}}{2004}]{2004ApJ...615....7M}
{Morales}, M.~F.,  \& {Hewitt}, J. 2004, \apj, 615, 7

\bibitem[\protect\citeauthoryear{{Myers} et~al.}{{Myers}
  et~al.}{2003}]{2003ApJ...591..575M}
{Myers}, S.~T., et~al. 2003, \apj, 591, 575

\bibitem[\protect\citeauthoryear{{Naoz} \& {Barkana}}{{Naoz} \&
  {Barkana}}{2005}]{2005MNRAS.362.1047N}
{Naoz}, S.,  \& {Barkana}, R. 2005, \mnras, 362, 1047

\bibitem[\protect\citeauthoryear{{Oh} \& {Mack}}{{Oh} \&
  {Mack}}{2003}]{2003MNRAS.346..871O}
{Oh}, S.~P.,  \& {Mack}, K.~J. 2003, \mnras, 346, 871

\bibitem[\protect\citeauthoryear{{Pen}, {Wu}, \& {Peterson}}{{Pen}
  et~al.}{2004}]{Pen_PAST}
{Pen}, U.~L., {Wu}, X.~P.,  \& {Peterson}, J. 2004, Ch.J.A.A., submitted

\bibitem[\protect\citeauthoryear{{Santos}, {Cooray}, \& {Knox}}{{Santos}
  et~al.}{2005}]{2005ApJ...625..575S}
{Santos}, M.~G., {Cooray}, A.,  \& {Knox}, L. 2005, \apj, 625, 575

\bibitem[\protect\citeauthoryear{{Scott} \& {Rees}}{{Scott} \&
  {Rees}}{1990}]{1990MNRAS.247..510S}
{Scott}, D.,  \& {Rees}, M.~J. 1990, \mnras, 247, 510

\bibitem[\protect\citeauthoryear{{Seljak} \& {Zaldarriaga}}{{Seljak} \&
  {Zaldarriaga}}{1996}]{1996ApJ...469..437S}
{Seljak}, U.,  \& {Zaldarriaga}, M. 1996, \apj, 469, 437

\bibitem[\protect\citeauthoryear{{Seo} \& {Eisenstein}}{{Seo} \&
  {Eisenstein}}{2003}]{2003ApJ...598..720S}
{Seo}, H.-J.,  \& {Eisenstein}, D.~J. 2003, \apj, 598, 720

\bibitem[\protect\citeauthoryear{{Spergel} et~al.}{{Spergel}
  et~al.}{2006}]{2006astro.ph..3449S}
{Spergel}, D.~N., et~al. 2006, ArXiv Astrophysics e-prints

\bibitem[\protect\citeauthoryear{{Spergel} et~al.}{{Spergel}
  et~al.}{2003}]{2003ApJS..148..175S}
{Spergel}, D.~N., et~al. 2003, \apjs, 148, 175

\bibitem[\protect\citeauthoryear{{Sunyaev} \& {Zeldovich}}{{Sunyaev} \&
  {Zeldovich}}{1972}]{1972A&A....20..189S}
{Sunyaev}, R.~A.,  \& {Zeldovich}, Y.~B. 1972, \aap, 20, 189

\bibitem[\protect\citeauthoryear{{Tegmark} et~al.}{{Tegmark}
  et~al.}{2004}]{2004PhRvD..69j3501T}
{Tegmark}, M., et~al. 2004, \prd, 69, 103501

\bibitem[\protect\citeauthoryear{{Tegmark}, {Taylor}, \& {Heavens}}{{Tegmark}
  et~al.}{1997}]{1997ApJ...480...22T}
{Tegmark}, M., {Taylor}, A.~N.,  \& {Heavens}, A.~F. 1997, \apj, 480, 22

\bibitem[\protect\citeauthoryear{{Tozzi} et~al.}{{Tozzi}
  et~al.}{2000}]{2000ApJ...528..597T}
{Tozzi}, P., {Madau}, P., {Meiksin}, A.,  \& {Rees}, M.~J. 2000, \apj, 528, 597

\bibitem[\protect\citeauthoryear{{Wang} et~al.}{{Wang}
  et~al.}{2004}]{2004PhRvD..70l3008W}
{Wang}, S., {Khoury}, J., {Haiman}, Z.,  \& {May}, M. 2004, \prd, 70, 123008

\bibitem[\protect\citeauthoryear{{Wang} et~al.}{{Wang}
  et~al.}{2006}]{Wang_Foreground_Subtraction}
{Wang}, X., {Tegmark}, M., {Santos}, M.,  \& {Knox}, L. 2006, \prd, submitted

\bibitem[\protect\citeauthoryear{{White} et~al.}{{White}
  et~al.}{1999}]{1999ApJ...514...12W}
{White}, M., {Carlstrom}, J.~E., {Dragovan}, M.,  \& {Holzapfel}, W.~L. 1999,
  \apj, 514, 12

\bibitem[\protect\citeauthoryear{{Wyithe} \& {Loeb}}{{Wyithe} \&
  {Loeb}}{2004}]{2004Natur.427..815W}
{Wyithe}, J.~S.~B.,  \& {Loeb}, A. 2004, \nat, 427, 815

\bibitem[\protect\citeauthoryear{{Wyithe}, {Loeb}, \& {Barnes}}{{Wyithe}
  et~al.}{2005}]{2005ApJ...634..715W}
{Wyithe}, J.~S.~B., {Loeb}, A.,  \& {Barnes}, D.~G. 2005, \apj, 634, 715

\bibitem[\protect\citeauthoryear{{Zaldarriaga}, {Furlanetto}, \&
  {Hernquist}}{{Zaldarriaga} et~al.}{2004}]{2004ApJ...608..622Z}
{Zaldarriaga}, M., {Furlanetto}, S.~R.,  \& {Hernquist}, L. 2004, \apj, 608,
  622

\end{thebibliography}

\clearpage
    \begin{deluxetable}{lc}
    \tablecaption
    {
        Fiducial Observation Parameters
        \label{tab_detector}
    }

    \tablehead{\colhead{Parameter} & \colhead{MWA-LFD} (MWA5000)}

    \startdata
    Array configuration, $\rho(r)$ (m$^{-2})$     & $\sim r^{-2}$\\
    Array diameter, D (m)                       & 1500  (3000) \\
    Bandwidth, B (MHz)                          & 32 \\
    Frequency resolution (kHz)                  & 8 \\
    Number of antennas, N                       & 500 (5000)

    \enddata
    \tablecomments
    {
        Array parameters for the MWA-LFD and MWA5000 reference experiments.
        \label{tab_fiducial}
    }
    \end{deluxetable}

\clearpage
    \begin{deluxetable}{lcccc}
    \tablecaption
    {
        Redshift Dependent Parameters
        \label{tab_redshift}
    }

    \tablehead{\colhead{} & \colhead{$z=8$} & \colhead{$z=10$}}

    \startdata
    Angular resolution ($^\circ$)      & 0.073 (0.036) & 0.089 (0.044)\\
    Antenna collecting area, dA (m$^2$)        & 14 & 18 \\
    Antenna response scale, $\Theta$ ($^\circ$) & 31 & 38 \\
    Frequency (MHz)                 & 158 & 129 \\
    System temperature, $T_{sys}$ (K)    & 440 & 690
    \enddata

    \tablecomments
    {
Characteristics of the fiducial observation that depend on frequency,
and thus on redshift, are listed along with their values at the $z=8$
and $z=10$ edges of the observation redshift range.  In the
calculations, the values of the fiducial parameters at intermediate
redshifts were linearly interpolated from the end points (except for
$\Theta$, see text). For the angular resolution, the MWA-LFD and
MWA5000 have different properties due to the larger size of the
MWA5000 (whose values are listed parenthetically).
    }

    \end{deluxetable}

\clearpage
    \begin{deluxetable}{lcc}
    \tablecaption{Model Power Spectrum Parametrization}

    \tablehead{ \colhead{Parameter} & \colhead{Symbol} & \colhead{Fiducial Value} }
    \startdata
    Normalization (at $k=0.05$ Mpc$^{-1}$)          & $A$ & 1.00 \\
    Matter density                                  & $\Omega_M$ & 0.30 \\
    Baryon density                                  & $\Omega_b$ & 0.05 \\
    Hubble constant                                 & $h$ & 0.70 \\
    Scalar spectral index (at $k=0.05$ Mpc$^{-1}$)  & $n_s$ & 1.00 \\
    Running index slope (at $k=0.05$ Mpc$^{-1}$)    & $\alpha_s \equiv d n_s / d \ln k$ & 0.00\\
    Residual foreground amplitude                   & $F$ & 0.00 \\
    Residual thermal offset                         & $N$ & 0.00
    \enddata

    \tablecomments{Eight parameters used to describe the measured power spectrum,
    and their values for the fiducial model. The first six rows give the basic parameters
    used to describe the matter power spectrum and the last two rows give the parameters to
    quantify the residual astrophysical foreground and residual thermal offset terms.
    \label{tab_cosmo}
    }

    \end{deluxetable}

\clearpage
    \begin{deluxetable}{l|ccccccccc}
    \tablecaption{Forecasted Uncertainties for the Model Parameters}

    \tablehead{ & \colhead{A} & \colhead{F} & \colhead{D} & \colhead{$\Omega_M$} & \colhead{$\Omega_b$} & \colhead{h} & \colhead{$n_s$} & \colhead{$\alpha_s$}}
    \startdata
            & 1.00 & 0.00 & 0.00 & 0.30 & 0.05 & 0.70 & 1.00 & 0.00 \\
    \hline
    \hline \\
    MWA-LFD     & 6.69 & 0.09 & 2e-3 & 0.21 & 0.07 & 0.88 & 0.2 & 0.04 \\
    MWA5000 & 0.45 & 2e-3 & 1e-4 & 0.01 & 6e-3 & 0.05 & 0.02 & 4e-3
    \\
    \hline
    \hline \\

    WMAP & 0.09 & --- & --- & 0.04 & 4e-3 & 0.04 & 0.03 & 0.02 \\
    SDSS & 0.10 & --- & --- & 0.05 & 2e-3 & 0.04 & 0.03 & ---

    \enddata

    \tablecomments{
    Forecasted 1-$\sigma$ uncertainties for the model parameters.  The first row gives the
    fiducial values for reference.  The MWA-LFD reference experiment
    does not significantly constrain cosmological parameters, while
    the MWA5000 does a reasonably good of job, particularly for
    $\Omega_M$ and the primordial power spectrum spectral index
    descriptors, $n_s$ and $\alpha_s$.  For comparison, the last
    two rows display approximate constraints for similar models from WMAP
    \citep[Their Table 10]{2003ApJS..148..175S} and SDSS
    \citep[Their Table 3]{2004PhRvD..69j3501T}.
    \label{tab_cov1}
    }

    \end{deluxetable}

    \begin{deluxetable}{l|cccccccc}
    \tablecaption{Covariance Matrix for MWA-LFD}

    \tablehead{ & \colhead{A} & \colhead{F} & \colhead{D} & \colhead{$\Omega_M$} & \colhead{$\Omega_b$} & \colhead{h} & \colhead{$n_s$} & \colhead{$\alpha_s$}}
    \startdata
    A & 1.00 & & & & & & & \\
    F & -0.09 & 1.00 & & & & & & \\
    D & -0.09 & 0.01 & 1.00 & & & & & \\
    $\Omega_M$ & -0.04 & -0.72 & -0.04 & 1.00 & & & & \\
    $\Omega_b$ & -0.87 & -0.24 & 0.04 & 0.49 & 1.00 & & & \\
    h & -0.97 & 0.24 & 0.11 & -0.20 & 0.72 & 1.00 & & \\
    $n_s$ & 0.95 & 0.04 & -0.15 & -0.12 & -0.82 & -0.92 & 1.00 & \\
    $\alpha_s$ & -0.41 & -0.37 & -0.15 & 0.63 & 0.69 & 0.23 & -0.32 & 1.00 \\
    \enddata

    \tablecomments{ Elements of the covariance matrix for the MWA-LFD
    calculated by taking the inverse of the Fisher matrix, $F$.  Each
    element has been normalized according to $c_{ab} / \sqrt{c_{aa}
    c_{bb}}$, where the diagonal elements are the squares of the 1-$\sigma$
    uncertainties given in Table \ref{tab_cov1}.
    \label{tab_cov2}}

    \end{deluxetable}

    \begin{deluxetable}{l|cccccccc}
    \tablecaption{Covariance Matrix for MWA5000}
    \tablehead{ & \colhead{A} & \colhead{F} & \colhead{D} & \colhead{$\Omega_M$} & \colhead{$\Omega_b$} & \colhead{h} & \colhead{$n_s$} & \colhead{$\alpha_s$}}
    \startdata
    A & 1.00 & & & & & & & \\
    F & 0.20 & 1.00 & & & & & & \\
    D & -0.07 & 0.00 & 1.00 & & & & & \\
    $\Omega_M$ & -0.40 & -0.60 & -0.02 & 1.00 & & & & \\
    $\Omega_b$ & -0.93 & -0.35 & 0.05 & 0.66 & 1.00 & & & \\
    h & -0.97 & -0.08 & 0.08 & 0.19 & 0.81 & 1.00 & & \\
    $n_s$ & 0.93 & 0.27 & -0.07 & -0.46 & -0.83 & -0.92 & 1.00 & \\
    $\alpha_s$ & -0.48 & -0.28 & -0.20 & 0.53 & 0.50 & 0.42 & -0.62 & 1.00 \\
    \enddata

    \tablecomments{Same as Table 3, but for the MWA5000. \label{tab_cov3}}

    \end{deluxetable}

\end{document}